\documentclass[a4paper]{spie}
\newif\ifpdf
\ifx\pdfoutput\undefined
\pdffalse 
\else
\pdfoutput=1 
\pdftrue
\fi

\ifpdf
\usepackage[pdftex]{graphicx}
\else
\usepackage{graphicx}
\fi

\usepackage{ae,amsmath,amssymb,exscale,subfigure}

\ifpdf
\DeclareGraphicsExtensions{.pdf, .jpg, .tif} \else
\DeclareGraphicsExtensions{.eps, .jpg} \fi 

\DeclareMathOperator{\abs}{abs}
\DeclareMathOperator{\cov}{cov}

\DeclareMathOperator{\var}{var}

\date{}

\begin{document}
\title{Nonlinear aspects of the EEG during sleep in children}
\author{Matthew J. Berryman\supit{a}, Scott W. Coussens\supit{b},
Yvonne Pamula\supit{c}, Declan Kennedy\supit{c}, Kurt Lushington\supit{d}, Cosma Shalizi\supit{e}, Andrew Allison\supit{a}, A. James Martin\supit{c}, David Saint\supit{b} and Derek Abbott\supit{a}
\skiplinehalf
\supit{a}Centre for Biomedical Engineering and
School of Electrical and Electronic Engineering,
The University of Adelaide, SA  5005, Australia.\\
\supit{b}Department of Physiology, School of Molecular and Biomedical Sciences. The University of Adelaide, SA 5005, Australia.\\
\supit{c}Department of Pulmonary Medicine, Women's and Children's Hospital, 72 King William Road, North Adelaide, SA 5006, Australia.\\
\supit{d}School of Psychology, University of South Australia, Adelaide, SA 5000, Australia.\\
\supit{e}Center for the Study of Complex Systems, 4485 Randall Laboratory, University of Michigan, Ann Arbor, MI 48109-1120, USA.}
\authorinfo{Send correspondence to Derek Abbott\\
E-mail: dabbott@eleceng.adelaide.edu.au, Telephone: +61 8 8303 5748}
\maketitle
\begin{abstract}
Electroencephalograph (EEG) analysis enables the neuronal behavior of a 
section of the brain to be examined. If the behavior is nonlinear then 
nonlinear tools can be used to glean information on brain behavior, and 
aid in the diagnosis of sleep abnormalities such as obstructive sleep apnea syndrome (OSAS). 
In this paper the sleep EEGs of a set of normal 
and mild OSAS children are evaluated for nonlinear behaviour. We consider how the behaviour of the brain changes with sleep stage and between normal and OSAS children.
\end{abstract}
\section{Introduction}\label{section:introduction}
\subsection{Overview}\label{section:inroduction:overview}
Electroencephalograph (EEG) analysis enables the neuronal behavior of a section of the brain to be examined~\cite{measurement}. As neurons themselves display nonlinear behavior, it is suspected that the overall behavior of groups of neurons is also nonlinear~\cite{CellAssemblies}. If the behavior is nonlinear, it allows the use of nonlinear statistics to describe the behavior of the brain~\cite{nonlinTSA}.

Measurement and analysis of the EEG is an integral part of the evaluation of sleep disorders in both adults and children. It is used in the classification of sleep architecture, a cyclic progression of sleep that is tightly controlled such that in adults a new cycle of REM (rapid eye movement) and Non-REM sleep occurs approximately every 90 minutes. A common respiratory sleep disorder is obstructive sleep apnea syndrome (OSAS). In OSAS the upper airway experiences repetitive periods of partial or complete occlusion during sleep. The disruption of sleep architecture by OSAS leads to well described daytime sequelae including reduced neurocognitive functioning, increased problematic behaviour, daytime sleepiness, impaired mood, and an increased risk of accidents. EEG parameters in combination with respiratory data are used to assess OSAS severity and these have been correlated with deficits in daytime functioning. EEG parameters can be derived through linear and nonlinear analyses. Evidence of linear and nonlinear brain activity has been demonstrated in adults~\cite{Stepien02,Das02} but very little research has been done in children; in particular there are conflicting results with different measures~\cite{Ferri02,Ferri03} and between children and young adults. It also remains to be demonstrated whether any observed nonlinearity reflects brain processes rather than nonlinearity of the amplifiers and other equipment used to collect the EEG data. Given the central role of sleep in neuronal development and plasticity it is imperative to establish in children the relationship of linearity and nonlinearity in brain behavior during sleep. This may also provide novel insights into the mechanism and effects of sleep architecture disruption caused by OSAS. In particular, it is important to test whether nonlinear parameters distinguish normal children from those with OSAS.
\section{Methods}
\subsection{Participants}
Thirteen children with a history of snoring and suspicion of OSAS participated in this study.  These children had been referred to a paediatric sleep disorders unit for evaluation of upper airway obstruction prior to adenotonsillectomy.  In addition, 13 non-snoring controls of a similar age range were also recruited into this study from friends of the snoring group or through newspaper advertisements.  All children underwent an overnight polysomnogram (PSG) to evaluate the degree of upper airway obstruction and to collect EEG data.  Other than a history of snoring in the former group, all children were otherwise healthy and not taking any medication that may influence EEG dynamics.  Informed consent was obtained from all parents of the children and where age appropriate, from the children themselves. This study was approved by the Women's and Children's Hospital Research Ethics Board
\subsection{Overnight polysomnography}
Overnight polysomnography (PSG) was conducted without sedation or sleep deprivation and began at each child's usual bedtime utilising standard protocols for children (American Thoracic Society, 1994).  A parent accompanied each child throughout the procedure.  The following standard parameters were measured and recorded continuously: electroencephalogram (EEG; C3-A2 or C4-A1), left and right electrooculogram (EOG), sub-mental and intercostal electromyogram (EMG) with skin surface electrodes, leg movements by piezoelectric motion detection, heart rate by electrocardiogram (ECG), oro-nasal airflow by thermistor and/or nasal pressure, respiratory movements of the chest and abdominal wall using uncalibrated respiratory inductive plethysmography (RIP), arterial oxygen saturation (SaO2) by pulse oximetry (Nellcor N200; three second averaging time) and transcutaneous CO2 (TcCO2) using a heated (314 K) transcutaneous electrode (TINA, Radiometer Pacific).  

All polysomnograms were analysed and scored manually by a sleep technician experienced and trained in analysing paediatric sleep studies.  Sleep stages were scored in 30-second epochs according to the standardised EEG, EOG and EMG criteria of Rechtschaffen and Kales~\cite{RechtschaffenKales} and included rapid eye movement (REM) sleep and the four stages (1-4) of non-rapid eye movement (NREM) sleep.  As stage 3 NREM sleep comprises only a small proportion of children's sleep it was combined with stage 4 NREM sleep and termed slow wave sleep (SWS) as is common practice.  Respiratory variables were scored according to standard guidelines recommended for paediatric sleep studies~\cite{Marcus92,BreathingStandards}. Obstructive apnoeas were defined as the absence of airflow associated with continued chest and abdominal wall movement for a duration of two or more respiratory cycles.  Obstructive hypopnoeas were defined as a 50-80\% reduction in the amplitude of the RIP and/or airflow signal associated with paradoxical chest/abdominal wall movement for a duration two or more respiratory cycles associated with either a 4\% oxygen desaturation and or EEG arousal. 
\subsection{EEG recordings}
The EEG data was recorded from the C3-A2 or C4-A1 position in the international 10-20 electrode placement system, with a reference point behind the mastoid. The signal was notch filtered to remove as much of the 50 Hz AC ripple as possible and amplified by an analog amplifier. The analog signal was sampled at 125 Hz and digitized using a linear digitizer. 
Artifact contamination of the EEG signals included extraneous signals from muscular movement~\cite{TransientArousals}, digitization noise, and also signals from the environment being picked up by unshielded EEG leads. Of particular concern is the nonlinear nature of filters and amplifiers used to process the analogue signal before digitization, as what is of interest is the nonlinearities in the underlying brain processes and not those of the equipment used.

\subsection{EEG data analysis}\label{section:introduction:methods}
As discussed by Schreiber and Schmitz~\cite{NonlinTest}, there are a number of methods for determining whether signals originate from nonlinear signals. There are a number of caveats with using these; not only do many of them require assumptions about the data, and have varying power of rejection of the null hypotheses of linearity. Of these, the best one overall seems to be the simple time reversibility test, which only requires assymetry of the data under visual inspection (that is, they have significant end effects). The time reversibility test computes a simple time reversal statistic on the data under test, and a set of linear surrogate data chosen carefully with the same general statistical properties~\cite{ImprovedSurrogate}. The particular statistic is given by
\begin{equation}
t_{r}=
\frac{\displaystyle \sum_{n=i_{d}+1}^{N}\left(x\left[n\right]-x\left[n-i_{d}\right]\right)^{3}}
       {\displaystyle \sum_{n=i_{d}+1}^{N}\left(x\left[n\right]-x\left[n-i_{d}\right]\right)^{2}},
\label{timeRevTest}
\end{equation}
where $i_{d}$ is a delay. Typically $i_{d}=1$ is sufficient, and is used in this paper.
As Schreiber and Schmitz note~\cite{NonlinTest}, this measure works best when there are only a few data sets with clear asymmetry under time reversal. Windows of length $10\,000$ from the EEG files that have significant end effects are used, and from this 19 sets of surrogate data of the same length are generated that have the same Fourier amplitudes and distribution; this provides a better null hypothesis than using a Gaussian linear process~\cite{ImprovedSurrogate}. If the value of $t_{r}$ for the original data is not the least out of the set of surrogate data, then the null hypothesis is rejected at the 95\% significance level, and hence shows nonlinearity.

It would be prudent to also use another measure of nonlinearity, and here the Higuchi fractal measure is used as it gives a number representative of the amount of nonlinearity in each individual window. The Higuchi fractal metric gives us a measure of the underlying nonlinear dynamics of a signal without trying to reconstruct a strange attractor~\cite{nonlinTSA,Higuchi,FDAnalysis}. The Higuchi measure provides a reliable measure of the fractal dimension when working with short time series segments, that is, those with sample length $N<125$~\cite{FDAnalysis}. It is also relatively insensitive to nonlinearities in noise or in amplification~\cite{FDAnalysis} so is useful for establishing that the nonlinear behavior comes from the underlying system, in this case the brain.
In the Higuchi fractal metric we calculate
\begin{equation}
L\left(k\right) = \displaystyle \sum_{m=0}^{k-1}\frac{N-1}{\lfloor\frac{N-m}{k}\rfloor k^{2}} \sum_{j=1}^{\lfloor\frac{N-m}{k}\rfloor}\abs\left[x\left(m+jk\right)-x\left(m+\left(j-1\right)k\right)\right],
\label{Lk}
\end{equation}
where $N=f_{s}\times8$s for an eight second window (hence $N=1000$ for $f_{s}=125$ Hz, or $N=2000$ for $f_{s}=250$ Hz), and $k=1,2,\ldots,2f_{s}$. Using a least squares fit of $y=\log\left(L\left(k\right)\right)$ against $x=\log k$ gives the Higuchi fractal measure $d_{H}$, 
\begin{equation}
d_{H}=-\frac{\cov\left(x,y\right)}{\var\left(x\right)}.
\label{d}
\end{equation}
The Higuchi fractal measure lies between 1 and 2 in theory; in practice because it is only an estimate it may lie slightly outside this range. The lower the value the ``less complex'' and linear the signal is. Higher values indicate signals that look more complicated and are nonlinear. 

To compare results between the two groups and between sleep stages, we use the unpaired Student's t-test, which first computes a t-value,
\begin{equation}
t=\frac{\bar{x}_{a}-\bar{x}_{b}}{\sqrt{\frac{s_{a}^{2}}{N}+\frac{s_{b}^{2}}{M}}}
\label{tval}
\end{equation}
where $\bar{x}$ denotes the mean and $s$ denotes the standard deviation for the two groups $a$ and $b$ with sizes $N$ and $M$ respectively.
The t-value from Eq.~\ref{tval} is then compared with a two-tailed Student's t-distribution of $N+M-2$ degrees of freedom to determine a significance level. 
This requires the data be approximately normal, and in the central limit theorem if we have enough data in both sets of data under consideration then the t-test can be safely used. We check this, and also checked the significance value by manually computing the probability distributions involved. 
\section{Results}
\subsection{Participants and PSG findings}
The thirteen snoring children comprised six males and had a mean age of 6.8 years (range 5.1 -- 8.7 years).  The control group also comprised of six males and had a mean age of 7.6 years (range 5.2 -- 10.9 years).  Overnight PSG analysis demonstrated that the snoring children had a higher number of obstructive apnoeas and hypopnoeas per hour of sleep (mean ($\pm$ SD) 0.6 (0.90)) than the non snoring control group (mean ($\pm$ SD) 0.03 (0.06)) and this difference was statistically different (P = 0.01, Mann Whitney U analysis).  However as the number of obstructive breathing events was less than one per hour of sleep in the snoring children, this is considered as having only very mild OSAS or primary snoring.  There was no significant difference in the amount of time that each group of children slept (7.84 hours for the snoring group vs 7.17 hours for the control group).  Similarly there was no significant difference in the amount of time spent in each sleep stage by both groups of children.  
\subsection{Verifying time reversal test}
\begin{figure}[hptb]
\centering
\subfigure[Time domain plot of the EEG data that the time reversal test indicates comes from the hypothesis of a linear model. Note the significant end effects -- the variance of the signal varies visibly throughout the plot. The time reversed signal is also shown.]{\includegraphics[width=8cm]{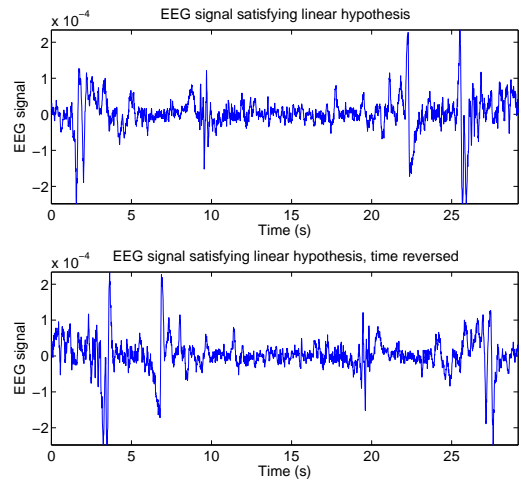}\label{linearEEG:a}}\\
\subfigure[The power spectrum of the EEG data fitting the linear hypothesis as shown in Subfigure (a). Surrogate data generated by a linear model with the same power spectra as the linear EEG has been generated and is plotted along with its power spectra to check they are identical.]{\includegraphics[width=8cm]{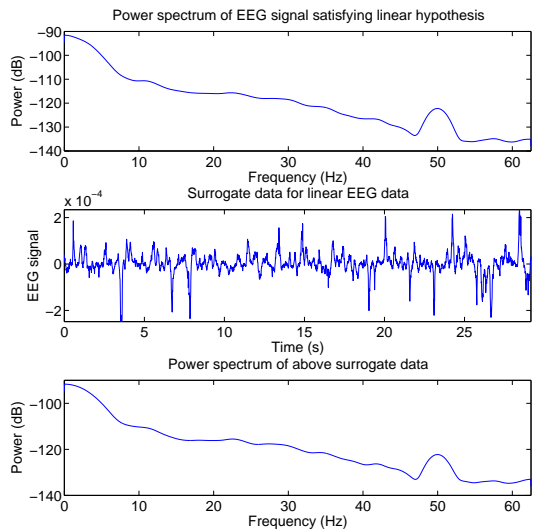}\label{linearEEG:b}}
\caption{The time plots of the EEG data fitting the linear hypothesis, its time reversal, and the surrogate data. Power spectra are also shown.} 
\label{linearEEG}
\end{figure}

\begin{figure}[hptb]
\centering
\subfigure[Time domain plot of the EEG data that the time reversal test indicates comes from the hypothesis of a nonlinear model. Note the significant end effects -- the variance of the signal varies visibly throughout the plot. The time reversed signal is also shown.]{\includegraphics[width=8cm]{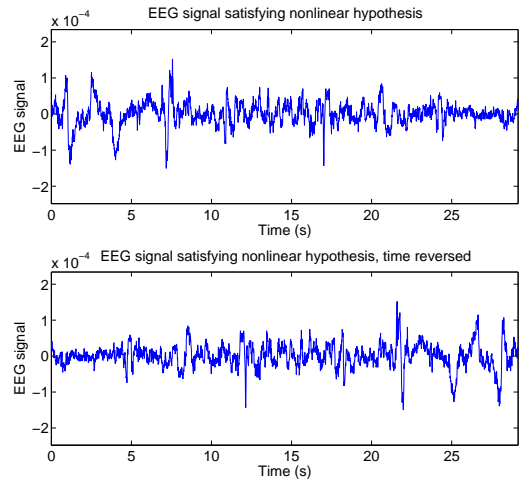}\label{nonlinearEEG:a}}\\
\subfigure[The power spectrum of the EEG data fitting the nonlinear hypothesis as shown in Subfigure (a). Surrogate data generated by a linear model with the same power spectra as the nonlinear EEG has been generated and is plotted along with its power spectra to check they are identical.]{\includegraphics[width=8cm]{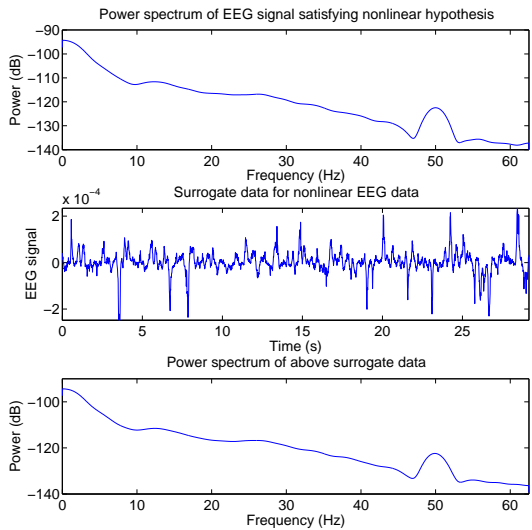}\label{nonlinearEEG:b}}
\caption{The time plots of the EEG data fitting the nonlinear hypothesis, its time reversal, and the surrogate data. Power spectra are also shown.} 
\label{nonlinearEEG}
\end{figure}

Visual verification that the data has significant end effects in order that the time reversal test can be used, and also that the surrogate data generated has the same power spectra as the original data was performed. Figures~\ref{linearEEG} and~\ref{nonlinearEEG} show that the data contains end effects, so the time reversal test can be safely used; furthermore they show that the surrogate method is correctly generating data with the same power spectra as the input data (the original time series) to the surrogate generation process.
\clearpage
\subsection{Time reversal test results}
For each sleep stage (including wake time after sleep onset) we calculated the percentage of the
 time that the time reversal test indicated significant nonlinear behavior (significant at the 95\% level). 
The results are shown for all the children combined and for the control and mild OSAS group separately, in Table~\ref{nonlin:single}. 
Table~\ref{nonlin:pairs} shows the differences (using the unpaired Student's t-test) in amount of nonlinear behaviour between different sleep states.
\begin{table}[hptb]
\centering
\caption{Mean percent of time (with standard deviation) the EEG was significantly nonlinear during different sleep/wake states computed from the time reversal statistic. Data are presented for clinical children, normal children, and for both groups combined.}
\begin{tabular}{llll}\hline
{\bf Sleep stage} & {\bf Patient group} & {\bf Control group} & {\bf Both groups}\\\hline
Non-REM 1 & 75.4 (12.7) & 68.0 (16.3) & 71.7 (14.8) \\
Non-REM 2 & 72.1 (8.7) & 69.5 (9.1) & 70.8 (8.8) \\
SWS & 68.7 (14.1) & 65.8 (12.6) & 67.3 (13.2) \\
Wake & 66.9 (9.9) & 52.8 (11.3) & 59.8 (12.7) \\
REM & 74.8 (6.7) & 67.5 (10.9) & 71.1 (9.6) \\
\hline
\end{tabular}
\label{nonlin:single}
\end{table}
\begin{table}[hptb]
\centering
\caption{Student's t-values comparing the amount of nonlinearity between sleep states, for the combined data set of control and mild OSAS children. Significance values: * = 95\%, **=99\%, ***=99.9\%.}
\begin{tabular}{lllll}\hline
{\bf Sleep stage} & REM & 1 Non-REM & 2 Non-REM & SWS \\\hline
Wake & 3.62*** & 3.10** & 3.63*** & 2.07*\\
REM & & 0.163 & -0.121 & -1.21\\
Non-REM 1 & & & 0.259 & 1.14\\
Non-REM 2 & & & & 1.14\\
\hline
\end{tabular}
\label{nonlin:pairs}
\end{table}
\subsection{Higuchi fractal results}
For each sleep stage (including wake after sleep onset) Higuchi fractal measures were calculated for all epochs in all subjects. These results are shown for the control and patient groups separately and for both groups combined in Table~\ref{higuchi:single}.
Table~\ref{higuchi:pairs} shows the differences (using the unpaired Student's t-test) for Higuchi fractal measures between different sleep states.

\begin{table}[hptb]
\centering
\caption{Higuchi fractal measures (mean $\pm$ SD) calculated for each 30 second epoch of data across
the group of clinical children, the group of normal children, and both groups combined. The Higuchi fractal measure here indicates that the data is generally linear across all sleep stages and groups. The values for the patient group are typically higher, although this is not significant.}
\begin{tabular}{llll}\hline
{\bf Sleep stage} & {\bf Patient group} & {\bf Control group} & {\bf Both groups}\\\hline
Non-REM 1 & 1.113 (0.084) & 1.079 (0.095) & 1.099 (0.090) \\
Non-REM 2 & 1.113 (0.052) & 1.097 (0.069) & 1.105 (0.061) \\
SWS & 1.099 (0.038) & 1.067 (0.081) & 1.083 (0.066) \\
Wake & 1.103 (0.140) & 1.095 (0.095) & 1.098 (0.116) \\
REM & 1.127 (0.041) & 1.115 (0.061) & 1.121 (0.051) \\
\hline
\end{tabular}
\label{higuchi:single}
\end{table}
\begin{table}[hptb]
\centering
\caption{This table shows the Student's t-values for sets of Higuchi fractal measures between sleep states, with the total set of data from both (control and patient) groups. Significance values: * = 95\%, **=99\%, ***=99.9\%.}
\begin{tabular}{lllll}\hline
{\bf Sleep stage} & REM & 1 Non-REM & 2 Non-REM & SWS \\\hline
Waking & 13.3*** & 0.200 & 4.53*** & -8.47***\\
REM & & -10.4*** & -17.2*** & -35.3***\\
1 Non-REM & & & -2.60** & 6.53***\\
2 Non-REM & & & & 21.9***\\
\hline
\end{tabular}
\label{higuchi:pairs}
\end{table}
\section{Discussion}
Comparing the same sleep states between patient and control groups reveals no significant difference in percent nonlinearity for any sleep states. Comparing Higuchi fractal measures however, we find a significant difference between patient and control groups for the Higuchi fractal measure in REM sleep only (t-value  9.45). This sleep state has been shown to be associated with learning~\cite{LocalSleepLearning,sleepmem}, and disrupting this sleep, as in OSAS children, affects learning~\cite{fMRIsleep}.

The typically low values of the Higuchi fractal measure, being close to one, confirm the general linear trend indicated by the time reversal test. It also reveals that the nonlinearity is due to underlying brain behaviour and not instrument noise. The amount of nonlinear brain behaviour is highest in NonREM stages 1 and 2 in addition to REM sleep, and lowest during wake (after sleep onset) and slow wave sleep. This is not a surprising finding for slow wave sleep, with predictable, low frequency waveforms present. It is somewhat surprising for wake after sleep onset, however it may represent simply the presence of linear muscle signals contaminating the EEG signal. 

Given that we have established nonlinearity in sleep stages of interest, in particular those associated with memory, it would make sense to use nonlinear measures to try and capture the brain behaviour. Linear measures (such as the often-used Fourier transform) should not be discounted however, since there is clear linearity throughout all sleep stages, and the signal may still be considered to be relatively stationary over local regions even when nonlinearity is present, as indicated by the low Higuchi fractal measures. Using the Higuchi fractal measure reveals a significant difference between control and patient groups in REM sleep, and this will be explored further in future work. It remains to be seen whether nonlinear measures are useful in classifying sleep stages, our work has highlighted the fact that the Higuchi fractal measure does not appear useful for this in children, who present difficulties  even for highly trained technicians in classifying sleep stages.
\section{Conclusions}
We have established some nonlinearity in the processes generating the EEG data using the time reversal and Higuchi tests, however there are considerable amounts of data that do not appear to be generated by nonlinear processes, in line with Stepie\'{n}~\cite{Stepien02}. Due to the significant changes in nonlinearity between sleep stages, and the Higuchi fractal measure, we can be certain that the nonlinearity process arises in the brain and not as a result of any nonlinear processes in the recording equipment. 

This work highlights the need to test for linearity before using nonlinear measures in evaluating EEG measure, in particular in distinguishing different brain states. Future work should focus on both linear and nonlinear measures for detecting local sleep events, such as apneas, as these may affect memory consolidation during sleep. The Higuchi fractal measure may be useful for this.
\bibliography{phd}
\bibliographystyle{spiebib}
\end{document}